\documentclass[8.5pt,twoside,twocolumn]{article}
\usepackage[super,sort&compress,comma]{natbib} 
\usepackage{mhchem}
\usepackage{times,mathptm}
\usepackage{sectsty}
\usepackage{balance} 
\usepackage{color}

\usepackage{graphicx} 
\usepackage{lastpage}
\usepackage[format=plain,justification=raggedright,singlelinecheck=false,font=small,labelfont=bf,labelsep=space]{caption} 
\usepackage{fancyhdr}
\usepackage{epic}
\begin{document}

\thispagestyle{plain}
\fancypagestyle{plain}{
\renewcommand{\headrulewidth}{1pt}}
\renewcommand{\thefootnote}{\fnsymbol{footnote}}
\renewcommand\footnoterule{\vspace*{1pt}%
\hrule width 3.4in height 0.4pt \vspace*{5pt}} 

\makeatletter 
\renewcommand\@biblabel[1]{#1}            
\renewcommand\@makefntext[1]%
{\noindent\makebox[0pt][r]{\@thefnmark\,}#1}
\makeatother 
\renewcommand{\figurename}{\small{Fig.}~}
\sectionfont{\large}
\subsectionfont{\normalsize} 

\fancyfoot{}
\fancyhead{}
\renewcommand{\headrulewidth}{1pt} 
\renewcommand{\footrulewidth}{1pt}
\setlength{\arrayrulewidth}{1pt}
\setlength{\columnsep}{6.5mm}
\setlength\bibsep{1pt}

\twocolumn[
  \begin{@twocolumnfalse}
\noindent\LARGE{\textbf{Switching dynamics in cholesteric blue phases}}
\vspace{0.6cm}

\noindent\large{\textbf{A. Tiribocchi$^{\ast}${$^{a}$}, G. Gonnella$^{a}$, D. Marenduzzo$^{b}$, and E. Orlandini\textit{$^{c}$}}}\vspace{0.5cm}

\noindent\textit{\small{\textbf{Received Xth XXXXXXXXXX 20XX, Accepted Xth XXXXXXXXX 20XX\newline
First published on the web Xth XXXXXXXXXX 200X}}}

\noindent \textbf{\small{DOI: 10.1039/b000000x}}
\vspace{0.6cm}

\noindent \normalsize{
Blue phases are networks of disclination lines, which occur in cholesteric liquid crystals near the transition to the isotropic phase. They have recently been used for the new generation of fast switching liquid crystal displays. Here we study numerically the steady states and switching hydrodynamics of blue phase I (BPI) and blue phase II (BPII) cells subjected to an electric field. When the field is on, there are three regimes: for very weak fields (and strong anchoring at the boundaries) the blue phases are almost unaffected, for intermediate fields the disclinations twist (for BPI) and unzip (for BPII), whereas for very large voltages the network dissolves in the bulk of the cell. Interestingly, we find that a BPII cell can recover its original structure when the field is switched off, whereas a BPI cell is found to be trapped more easily into metastable configurations. The kinetic pathways followed during switching on and off entails dramatic reorganisation of the discli
 nation networks. We also discuss the effect of changing the director field anchoring at the boundary planes and of varying the direction of the applied field.}

\vspace{0.5cm}
 \end{@twocolumnfalse}
  ]

\section{Introduction}

\footnotetext{\textit{$^{a}$~Dipartimento di Fisica and Sezione INFN di Bari, Universita' di Bari, I-70126 Bari, Italy}}
\footnotetext{\textit{$^{b}$~SUPA, School of Physics and Astronomy, University of Edinburgh, Mayfield Road, Edinburgh EH9 3JZ, UK}}
\footnotetext{\textit{$^{c}$~Dipartimento di Fisica and Sezione INFN di Padova, Universita' di Padova, Via Marzolo 8, 35131 Padova, Italy}}

In many cholesteric liquid crystals (CLC) the transition between the
regular cholesteric phase and the isotropic phase occurs through a
cascade of intermediate phases known as ``blue phases''
(BP)~\cite{mermin}. Blue phases occur because the locally
energetically favoured double twisting (i.e. local twisting in two
directions, as opposed to the single twist of the cholesteric) is
incompatible with the requirement of continuity~\cite{mermin}. 
This results in the formation of networks of defects that
separate isolated stable regions. The network can lead to quite
intricate structures that are periodic at length scales comparable to
the wavelength of light -- this fact is at the origin of their name 
(indeed blue phases can be of many different colours).

For example the BPI and BPII phases have a 3D cubic orientational
order with lattice periods of several hundred nanometres, and they
exhibit selective Bragg reflections in the range of visible light.
For this reason they  have interesting applications in fast light
modulators~\cite{dim89,heppke91} or tunable photonic
crystals~\cite{photonic}. Blue phases are in general
thermodynamically stable over a narrow temperature range but
recently their stabilization over a range of more than 60 K,
including room temperature, have been
shown~\cite{psbp,hisakado_2004,Coles_Pivnenko:2005:Nature, Karatairi_et_al:2010:Phys_Rev_E}. This has opened up new avenues in
liquid crystal technologies culminating in the recent fabrication of
a blue-phase display device with very fast switching and response
times~\cite{samsung}.

Blue phases are fully 3D structures and the understanding of even
their statics is quite a difficult task to achieve that requires
very refined techniques and important computational efforts. For
these reasons  theoretical investigations of blue phases in presence
of an electric field have been essentially restricted to the case of
infinite samples. Furthermore, typically analytical and semi-analytical 
theories have focused on the case in which the electric field is
infinitesimally small~\cite{cladis1,cladis2}. 
This is in contrast with the experimental characterization of the known blue 
phases in electric and magnetic fields~\cite{ter95,lasers,ef_exp1,ef_exp2}.

Therefore here we aim to investigate the switching dynamics of
cholesteric blue phases in an electric field, by numerically solving
the hydrodynamic equations of motion for BPI and BPII.
The choice of a realistic boundary condition is far from trivial in our system, and in this work we will compare and discuss different possibilities. We first focus on the cases in which the director field is fixed at the top and bottom boundaries ($z=0$ and $z=L$ respectively, where $L$ is the sample size), to its stable structure in the absence of a field -- this may in practice be achieved by pinning due to impurities or via surface ``memory'' effects~\footnote{Note that similar fixed boundaries have been used and have proved very useful to get physical insights into, for instance, permeation flows in cholesterics~\cite{perm1,perm2,perm3,degennes}, where they lead to better comparison with experiments as opposed to other boundary conditions. We will discuss boundary conditions and their roles more in detail when commenting the results later on.}.
We consider the switching dynamics both with an electric (or magnetic) field can be
switched on either along the $[0,0,1]$ $z$ direction or along the $[1,1,1]$
diagonal of the unit cell (Section IIIB). Since in practice the
orientation of the director field at the boundaries can be
controlled and varied (for example by rubbing the plates or treating
them chemically),  we then investigate the effect that different
alignments conditions at the boundaries may have on the structure in the absence and presence of a field (Section IIIC).
With respect to previous work on blue phases in electric fields~\cite{ef_theor1,ef_theor2,gareth1,fukuda2,fukuda1,oliver,domaingrowth}, our work considers hydrodynamic interactions, and focuses on switching on and off dynamics in a confined as opposed to a bulk system. This case should be relevant to devices which are typically in the $\mu$m range.

{ A general picture that emerges out of this study is
that} the switching is intimately interconnected with the dynamics
of the disclination network constituting these phases.  For example
we show that, under a small field, the disclination network of BPI
twists and that of BPII stretches and unzips at the center. Above a
given threshold of the voltage, in both cases the network breaks
down. The disclinations then pin to the boundaries 
(we restrict ourselves to the case in
which the BP texture is anchored at the top and bottom plates). For BPII we
observe that the breaking and pinning dynamics involve the creation
of a disclination loop in the middle of the sample. This indeed
allows the rotation of the disclination pattern at the boundaries
preserving its continuity. As the field is switched
{off}, the structure goes back to the original state
or to a cholesteric structure according to the steady state attained
with the field on. Measurements of the {time
evolution of the free-energy of the system allow to characterize the
intermediate states reached by the system during  the switching on
and off cycles.}

\section{Models and methods}

Equilibrium properties of blue phases are described by a Landau-de Gennes free energy density
written in terms of the tensor order parameter $Q_{\alpha\beta}$~\cite{degennes,chandrasekar,mermin}.
This comprises a bulk term
\begin{eqnarray}
f_{b}  =  \frac {A_0}{2} (1 - \frac {\gamma} {3}) Q_{\alpha \beta}^2 & - & 
          \frac {A_0 \gamma}{3} Q_{\alpha \beta}Q_{\beta
          \gamma}Q_{\gamma \alpha} \\ \nonumber
& + &  \frac {A_0 \gamma}{4} (Q_{\alpha \beta}^2)^2,
\label{eqBulkFree}
\end{eqnarray}
and a distortion term \cite{mermin}
\begin{equation}
f_{d} = \frac{K}{2} (\partial_\beta Q_{\alpha \beta})^2
+ \frac{K}{2} (\epsilon_{\alpha \zeta\delta }
\partial_{\zeta}Q_{\delta\beta} + 2q_0Q_{\alpha \beta})^2, 
\end{equation}
where $K$ is the elastic constant and the pitch of the cholesteric liquid crystal is
given by $p\equiv 2\pi/q_0$. The tensor $\epsilon_{\alpha \zeta\delta}$ is  the Levi-Civita
antisymmetric third-rank tensor, $A_0$ is a constant and $\gamma$ controls the magnitude of
order (it plays the role of an effective temperature for thermotropic liquid crystals).
{
Notice that the magnitude of order, $q(\vec r)$ (which lies between $0$ and $1$), is $3/2$ times the largest eigenvalue of ${\bf Q}$. }

{The anchoring of the director field on the boundary surfaces
to a chosen director $\hat{n}^0$ is ensured by adding the surface term
\begin{eqnarray}
f_s & = & \frac{1}{2}W_0 (Q_{\alpha \beta}-Q_{\alpha \beta}^0)^2\\
Q_{\alpha \beta}^0 & = & S_0 (n_{\alpha}^0n_{\beta}^0-\delta_{\alpha\beta}/3)
\end{eqnarray}
where the parameter $W_0$ controls the anchoring strength, while $S_0$ determines the degree of the surface order. If the surface order is equal to the bulk order, $S_0$ should be taken equal to $\bar{q}=1/2$ (the value of $q(\vec r)$ in the bulk, away from disclinations). In the case of strong anchoring $W_0$ is practically set to infinity, which means that the order parameter ${\bf Q}$ is fixed on the surfaces -- this is how we imposed the anchoring in our simulations.
 
The last contribution to the free-energy is given by $-(\epsilon_a/12\pi)E_{\alpha}Q_{\alpha\beta}E_{\beta}$, and 
describes the interactions with an external electric field $E$, where $\epsilon_a>0$ is the dielectric anisotropy of the liquid
crystal. The reduced potential is $\Delta V_z=E_z L_z$, where $E_z$ is the component of the electric
field along the $x$-direction and $L_z$ is the size
of the cell in the same direction. Similar expressions can be written for the $x$ and the $y$ components. 
The time evolution of the system is then governed by the equation of motion for the order
parameter~\cite{beris,perm3},
\begin{equation}
(\partial_t+{\vec u}\cdot{\bf \nabla}){\bf Q}-{\bf S}({\bf W},{\bf
  Q})= \Gamma {\bf H}
\label{Qevolution}
\end{equation}
where $\Gamma$ is a collective rotational diffusion constant.
The first term on the left-hand side of Eq. (\ref{Qevolution})
is the material derivative describing the usual time dependence of a
quantity advected by a fluid with velocity ${\vec u}$. Since for rod-like
molecules the order parameter distribution can be both rotated and stretched by
flow gradients~\cite{beris} a second term must be added to the material derivative:
\begin{eqnarray}\label{S_definition}
{\bf S}({\bf W},{\bf Q})
& = &(\xi{\bf D}+{\bf \omega})({\bf Q}+{\bf I}/3)+ ({\bf Q}+
{\bf I}/3)(\xi{\bf D}-{\bf \omega}) \nonumber \\
& - & 2\xi({\bf Q}+{\bf I}/3){\mbox{Tr}}({\bf Q}{\bf W})
\end{eqnarray}
where Tr denotes the tensorial trace, while
${\bf D}=({\bf W}+{\bf W}^T)/2$ and ${\bf \omega}=({\bf W}-{\bf W}^T)/2$
are the symmetric and the anti-symmetric part of the
velocity gradient tensor $W_{\alpha\beta}=\partial_\beta u_\alpha$.
The constant $\xi$ depends on the aspect ratio of the molecules forming a
given liquid crystal. This parameter controls whether  the director field is flow aligning
in shear flow ($\xi > \frac{3\bar{q}}{2+\bar{q}}$), creating a stable response, as opposed to flow tumbling, which
gives an unstable response ($\xi < \frac{3\bar{q}}{2+\bar{q}}$).}

The molecular field ${\bf H}$, which provides the driving motion towards the minimum of the
free energy, is given by
\begin{equation}
{\bf H}= -{\delta {\cal F} \over \delta {\bf Q}}+({\bf
    I}/3) Tr{\delta {\cal F} \over \delta {\bf Q}}.
\label{molecularfield}
\end{equation}
The fluid velocity, $\vec u$, obeys the continuity equation and the
Navier-Stokes equation, which
{in the incompressibile fluid limit} reads

\begin{eqnarray}\label{navierstokes}
\rho(\partial_t+ u_\beta \partial_\beta)
u_\alpha = \partial_\beta (\Pi_{\alpha\beta})+
\eta \partial_\beta(\partial_\alpha
u_\beta + \partial_\beta u_\alpha).
\end{eqnarray}
with a stress tensor generalized to describe the flow of cholesteric liquid crystals,
and equal to
\begin{eqnarray}
\Pi_{\alpha\beta}= &-&P_0 \delta_{\alpha \beta} +2\xi
(Q_{\alpha\beta}+{1\over 3}\delta_{\alpha\beta})Q_{\gamma\epsilon}
H_{\gamma\epsilon}\\\nonumber
&-&\xi H_{\alpha\gamma}(Q_{\gamma\beta}+{1\over
  3}\delta_{\gamma\beta})-\xi (Q_{\alpha\gamma}+{1\over
  3}\delta_{\alpha\gamma})H_{\gamma\beta}\\ \nonumber
&-&\partial_\alpha Q_{\gamma\nu} {\delta
{\cal F}\over \delta\partial_\beta Q_{\gamma\nu}}
+Q_{\alpha \gamma} H_{\gamma \beta} -H_{\alpha
 \gamma}Q_{\gamma \beta}
\label{BEstress}
\end{eqnarray}
{where $\rho$ is the fluid density and $\eta$ an
isotropic viscosity. 
$P_0$ is a constant in the simulations
reported here.}

{Notice that, unless the flow
field is zero ($\vec{u}=0$), the dynamics of the order parameter are
not purely relaxational. Conversely, the order parameter field
affects the dynamics of the flow field through the stress tensor,
which  depends on the
molecular field ${\bf H}$ and on ${\bf Q}$.
This is the back-flow coupling. In our simulations, the coupling to
hydrodynamics via a non-trivial pressure tensor can be switched off,
essentially by imposing a constant zero velocity profile. In this
way the effects of flow and backflow may be easily tested.}

{ To solve these equations we use a 3D hybrid lattice
Boltzmann algorithm that consists in solving Eq.~(\ref{Qevolution}) via
a finite difference predictor corrector algorithm while the
integration of the Navier-Stokes equation~(\ref{navierstokes}) is
taken care by a standard Lattice Boltzmann (LB) approach. The hybrid
method has the advantage that involves consistently smaller memory
requirements with respect to the full LB
counterpart~\cite{colin,proc} and its efficiency has been indeed
already tested for binary fluids~\cite{gonnella}, nematic devices~\cite{adri} and active nematic
liquid crystals~\cite{Marenduzzo_et_al:2007:Phys_Rev_E,Tiribocchi,softmatter}. } 

{In our simulations we fixed $\Gamma = 0.3$, $\xi =
0.7$ and $\eta = 1.333$ as suggested by
previous experience  on numerical investigation of blue
phases~\cite{alex}. Typical simulations reported here required about 2000 single processor cpu hours. Most of our runs were parallel (MPI architecture) with 8-16 nodes.} 

{
To locate the defects in the simulation domain, we look at the local behaviour of the order,
$ q(\vec{r})$: namely
when $q(\vec{r})$ falls below a predetermined threshold, we identifiy  that lattice point as
belonging to a disclination. This prescription is very easy to implement and allows
a determination of the disclinations that turned out to be as accurate as the more
complicated procedure based on the degree of biaxiality~\cite{Biscari_et_al:2006:Phys_Rev_E}.
For our study we have chosen a threshold of $q=0.19$ that corresponds to the 
$60\%$ of the largest eigenvalue of ${\bf Q}$ at the steady state.
Clearly a variation of the threshold value generally leads to a variation in the 
thickness of the rendering of the disclinations in the figures.}

In our calculations the size of a cubic unit cell has been set equal to $32\times 32\times 32$ in simulation units.
Note that the size of the unit cell which minimises the free energy is in general different from the cholesteric pitch -- typically it is larger, and this size change has been referred to as ``redshift'' in the literature~\cite{alex,alex2,gareth2}. In our simulations we have kept the redshift constant to the values suggested in Ref.~\cite{alex}, and valid for zero field. We note that in principle the electric field modifies the optimal redshift in equilibrium~\cite{oliver}. While this is important to obtain an accurate phase diagram for blue phases in the bulk, it does not qualitatively affect either the switching dynamics or the behaviour of a confined blue phase in an electric field (we have verified this by running our simulations with no redshift, both the succession of phases in Fig.~1 and the switching dynamics are unaffected).

\section{Results}

\subsection{Steady state in an electric field}

The equilibrium structures in the absence of an electric field (Fig.~1a and Fig.~1d) were obtained by relaxing the free energy to its minimum value by solving Eq. (\ref{Qevolution}) numerically. Periodic boundary conditions were employed along the $x$ and $y$ axes, whereas on the $z=0$, $z=L$ planes we fixed the director field as in the bulk equilibrated blue phase configuration. 
Before the network disrupts, also the experiments in Ref.~\cite{psbp} suggest that the texture at the boundaries does not modify so that our assumption should be appropriate. In practice, as commonly assumed in the theory of permeation in cholesterics~\cite{perm1,perm2,perm3}, fixed boundary conditions may be due to pinning of disclinations via surface impurities, or by surface ``memory'' effects (see below for results with different boundaries and a more thorough discussion of their roles).

When we apply an electric field we initially consider a system with width $L$ equal to the periodicity of the BP structures along z. Initial configurations were set according to the approximate solutions in Ref.~\cite{mermin}. Note that the
periodicity of the ground state is related to the blue phase pitch via a non-trivial constant (also known as ``redshift'') which we have taken to be equal to the one resulting from the semi-analytical minimisation of the free energy presented in Ref.~\cite{softmatter,gareth2,alex,ef_theor1}. The structure we get
then only depends on the liquid crystal chirality $\kappa$,
reduced temperature $\tau$ and on ${\cal E}$ which takes into account electric
field effects. These are related to our parameters as
{
\begin{eqnarray}
\kappa &=& \sqrt{\frac{108 K q_0^2}{A_0 \gamma}}\\
\tau &=& \frac{27(1-\gamma/3)}{\gamma}+ \kappa^2\\
{\cal E}^2&=&(27 \varepsilon_a E_\alpha E_\alpha)/(32 \pi A_0 \gamma).
\end{eqnarray}
}
The static phase diagram at ${\cal E}=0$ is in a better agreement with
experiments with respect to earlier calculations based on
semi-analytical approximations and is given in
Refs.~\cite{gareth2,alex}, where it is shown that BPI and BPII appear in order of increasing values of $\kappa$ as found experimentally.

{We have chosen $A_0$, $K$ and $\gamma$ in order to be in the appropriate
region of the phase diagram. We have set $A_0=0.0034$, $K=0.005$, $\gamma=3.775$
for BPI, and $A_0=0.00212$, $K=0.005$, $\gamma=2.9$ for BPII. These may correspond, for instance, to a blue phase with lattice constant equal to 400 nm, with a rotational viscosity equal to 1 Poise and (Frank) elastic constants equal to $\sim 10$ pN and $\sim 16$ pN for BPI and BPII respectively, which are in the range of typical liquid crystalline materials. Accordingly, one may calculate that one space and time LB unit correspond to 0.0125 $\mu$m and 0.0013 $\mu$s for BPI, and to 0.0125 $\mu$m and 0.0017$\mu$s for BPII.}
Furthermore, in order to map electric field related quantities to real units, the best avenue is to use the previously defined dimensionless number ${\cal E}$ and compare simulation and experimental values. An electric field of 1 V/$\mu$m, with a dielectric anisotropy $\sim$ 10, by assuming $\frac{27}{2 A_0\gamma} \sim 2-5 \times 10^{-6} {\rm J}^{-1}\, {\rm m}^{3}$~\cite{mermin}, would lead, for instance, to ${\cal E}\sim$0.01 whereas a voltage of 0.1 in Fig. 1 corresponds in our simulations to ${\cal E}\sim$0.1 ($\sim$0.09 for BPI and $\sim$0.13 for BPII).

In Fig.~1 we show how the steady state, after the field is switched
on, changes as a function of the {voltage applied along the $z$ direction}.
 In the case of BPII (top row), for
small electric field the steady state is indistinguishable from the
starting configuration (Fig.~1a), while above a threshold (a voltage
$\Delta V\simeq 0.175$ in simulation units with the parameters we
chose, see caption to Fig.~1), the network unzips at the centre
(Fig.~1b). For a voltage larger than about 0.3, the disclinations
reconstruct at the surface to form an array of straight parallel
lines (Fig.~1c). In the case of BPI (bottom row), the zero-field
structure (Fig.~1d), is unstable after a first threshold (0.275) and
becomes twisted in steady state with a voltage on (Fig.~1e). If the
field exceeds a second threshold (about 0.425 in
the case of Fig.~1), the network again breaks and the disclinations
pin to the boundaries (Fig.~1f).

\begin{figure*}
\centerline{\includegraphics[width=20.cm]{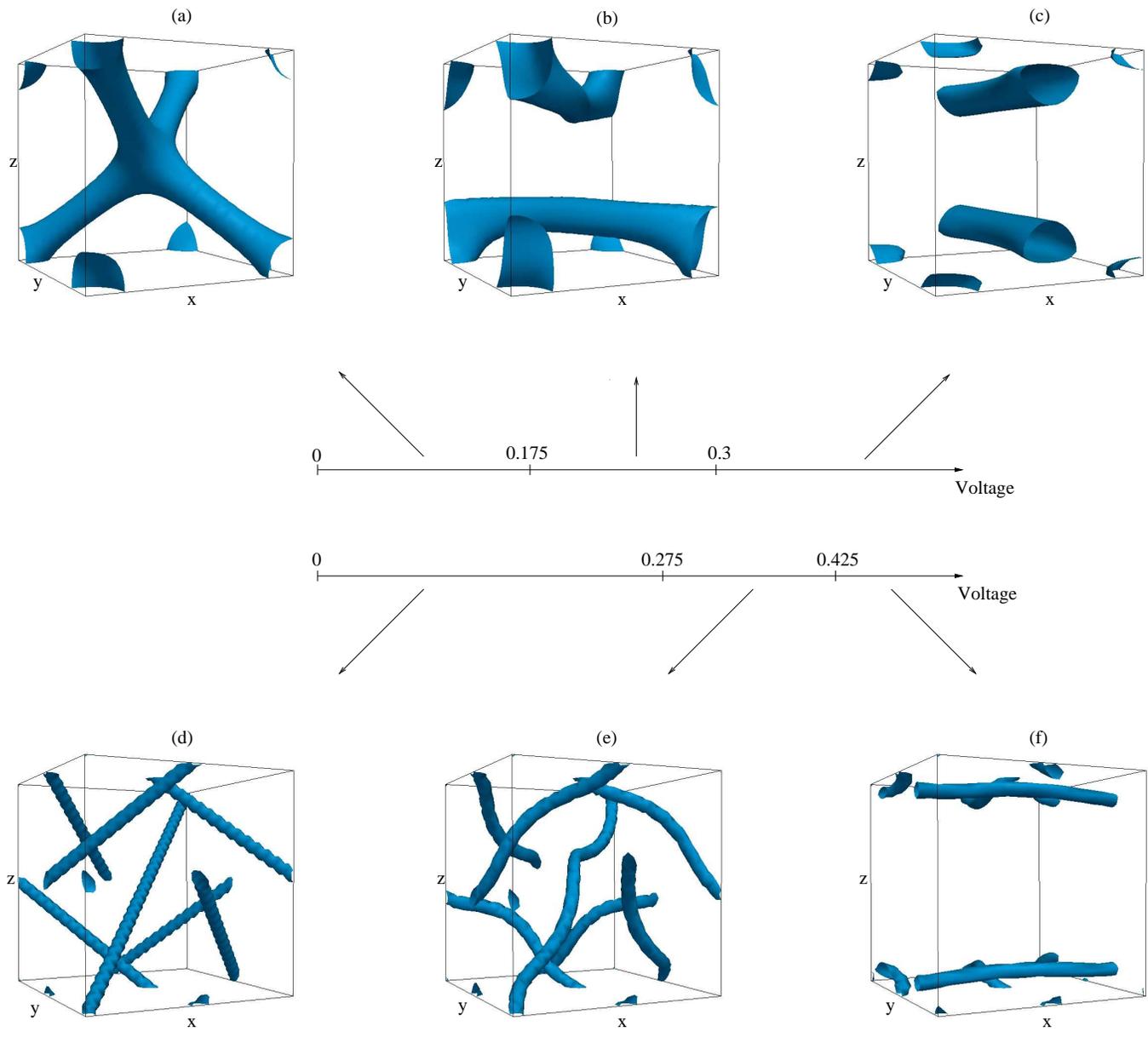}}
\caption{Steady state disclination networks for BPI and BPII under a varying field, and strong
anchoring of the zero-field structure on the surface. BPII structures are shown in the first row,
BPI are in the bottom row. The simulation domain has size $32\times 32\times 32$.
The range of stability in voltage (simulation units) of each of the
networks shown in the snapshots is also indicated. Snapshots correspond to steady states
with $\Delta V$ equal to 0 (a,d), 0.3 (b), 0.4 (c); 0.375 (e), 0.5 (f).}
\label{fig. 1}
\end{figure*}

The difference in the optical signal transmitted by a  
liquid crystalline sample between cross polarizers with
and without an applied field depends on the difference in the
director field profile. Fig.~2 shows the director field on a cut of
the sample parallel to the $xy$ plane at $z=L/2$, 
with different applied voltages. The double twist regions
separating the disclination lines - typical of blue phases in
equilibrium - are evident from the figure.

\begin{figure*}
\centerline{\includegraphics[width=20.cm]{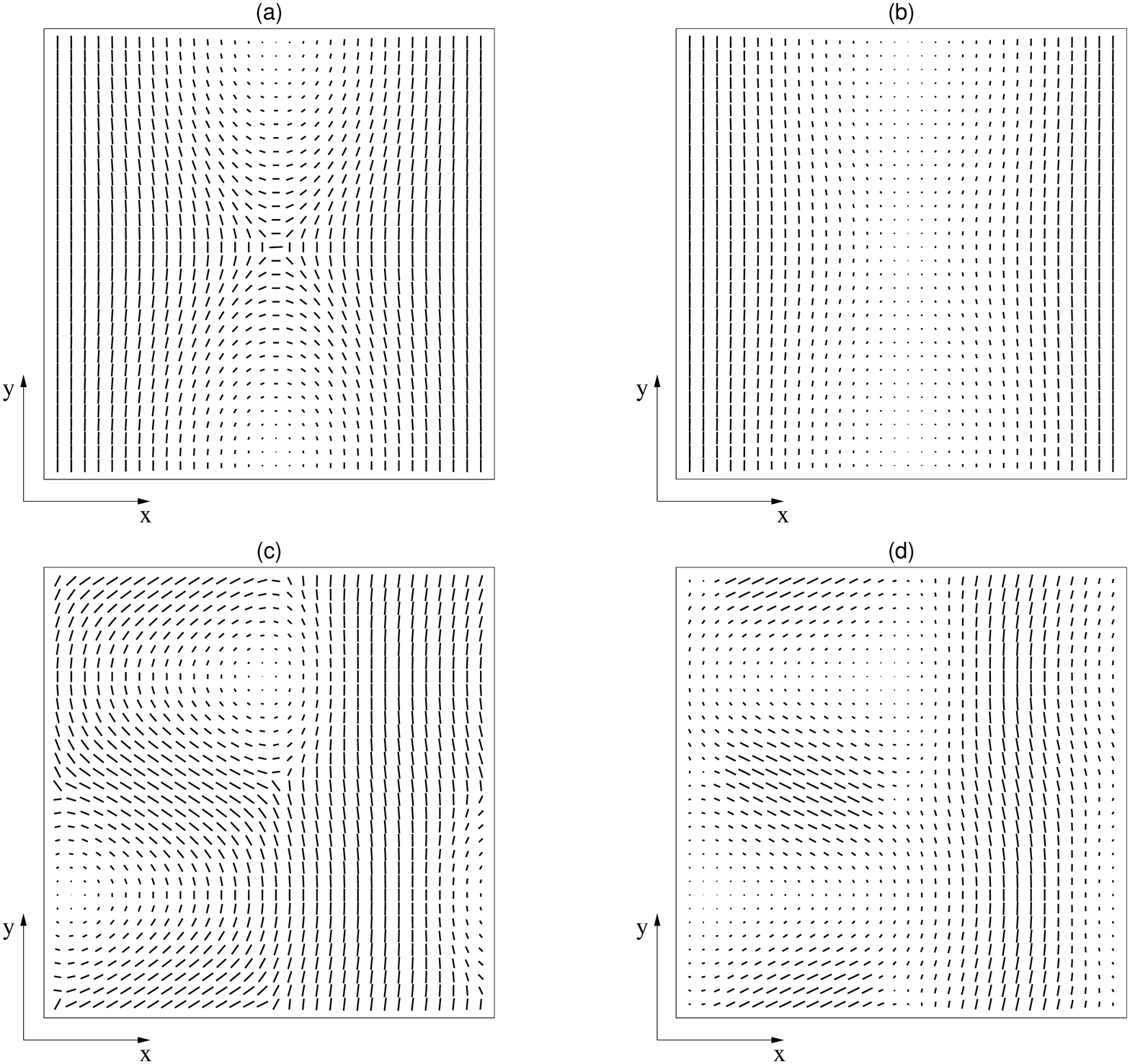}}
\caption{ Director field profiles on the $xy$ plane taken at $z=L/2$. Panels (a) and (b)
correspond to the steady state disclination networks of Fig~1a and 1b (BPII). Panels (c) and (d) correspond to Fig.~1d and Fig.~1e (BPI).}
\label{fig2}
\end{figure*}

\subsection{Switching dynamics of blue phase devices}

We now discuss the switching on and off dynamics of a blue phase
device -- focussing on the case of strong anchoring
{(i.e. $W_0 >>1$)}. Fig.~3 (snapshots $t_0-t_5$) shows the
response of a BPII cell to an electric field along the $\hat{z}$
direction. The field is strong enough to pin the disclinations to
the surface (see the preceding Section and the discussion of
Fig.~1). The disclinations unbind at the centre of the cell soon
after the application of the electric field. These then  join to the
defect at the boundaries, to reconstruct later on via the formation
of a loop in the bulk of the sample. Such a remodelling of the
defect networks leads to a rotation of the surface disclinations,
which end up straight and parallel to each other. 

It is interesting to ask what happens to this highly distorted BP
cell once the electric field is removed, {especially in the high field regime on which we focus later on.}
Do we get back to the
initial BPII configuration, or does the device get stuck into a
different metastable configuration?  To address these questions, we
have simulated the switching off dynamics starting from the
configuration shown in Fig.~3, snapshot $t_5$. When the electric field
is switched off to zero, the defects depin and meet in the bulk
(snapshots $t_6-t_7$). There, they join with the surface
defects due to the anchoring to form a fully connected structure
with the disclinations closeby in the center of the cell (snapshot
$t_8$). These then tilt and twist (snapshots $t_9-t_{10}$), to
finally reform the original structure (snapshot $t_{11}$). Therefore
the BPII cell has the potential to be used as an electrically
switchable device, provided at least that the original network of
defects is pinned to the surface. 

Fig.~4 shows the results of an analogous electric field cycle for a
BPI cell -- note that this is the BP used in the experimentally
constructed device in Ref.~\cite{Coles_Pivnenko:2005:Nature}.
Our results suggest that switching in this disclination network 
both entails a qualitatively different pathway and leads to a distinct steady 
state. During switching on ($t_0-t_5$), the BPI
disclinations transiently twist up, as in the steady states shown in
Fig.~1. When the twisted defects get close enough, they recombine
and once again pin to the boundaries. After the field is switched
off ($t_6-t_{11}$), different disclinations depin and follow
different dynamics: some migrate back to the centre, while
others rejoin to neighbouring ones to form either arcs connected to
the same surface, or lines spanning the whole device. In this case,
then, the network gets stuck ``en route" into a metastable state,
which is distinct from both the zero and the high field BPI
configurations. Despite this, {it is still possible} to use this
BPI cell as a device, which switches between the configurations in
$t_5$ and $t_{11}$ under the action of an applied field. We have done
systematic simulations of several cycles and we have observed that 
there is a reversible cycle between the field induced state $t_5$ and the 
metastable state $t_{11}$, but the system can not get back to the original
BPI defect structure as at $t_0$. Therefore the structure is permanently deformed.
It may be
interesting to systematically scan the chirality--temperature phase
space to investigate whether our finding that BPI devices do not
switch back to their original, zero field, stable state is robust
and generic, or depends on the exact position of the starting state
in the phase diagram. {To answer this question, we have performed a simulation
similar to the one described in Fig. 4 (i.e. field cycle for a BPI phase)
but starting from a different point of the phase diagram.} 

\begin{figure*}
\centerline{\includegraphics[width=18.cm]{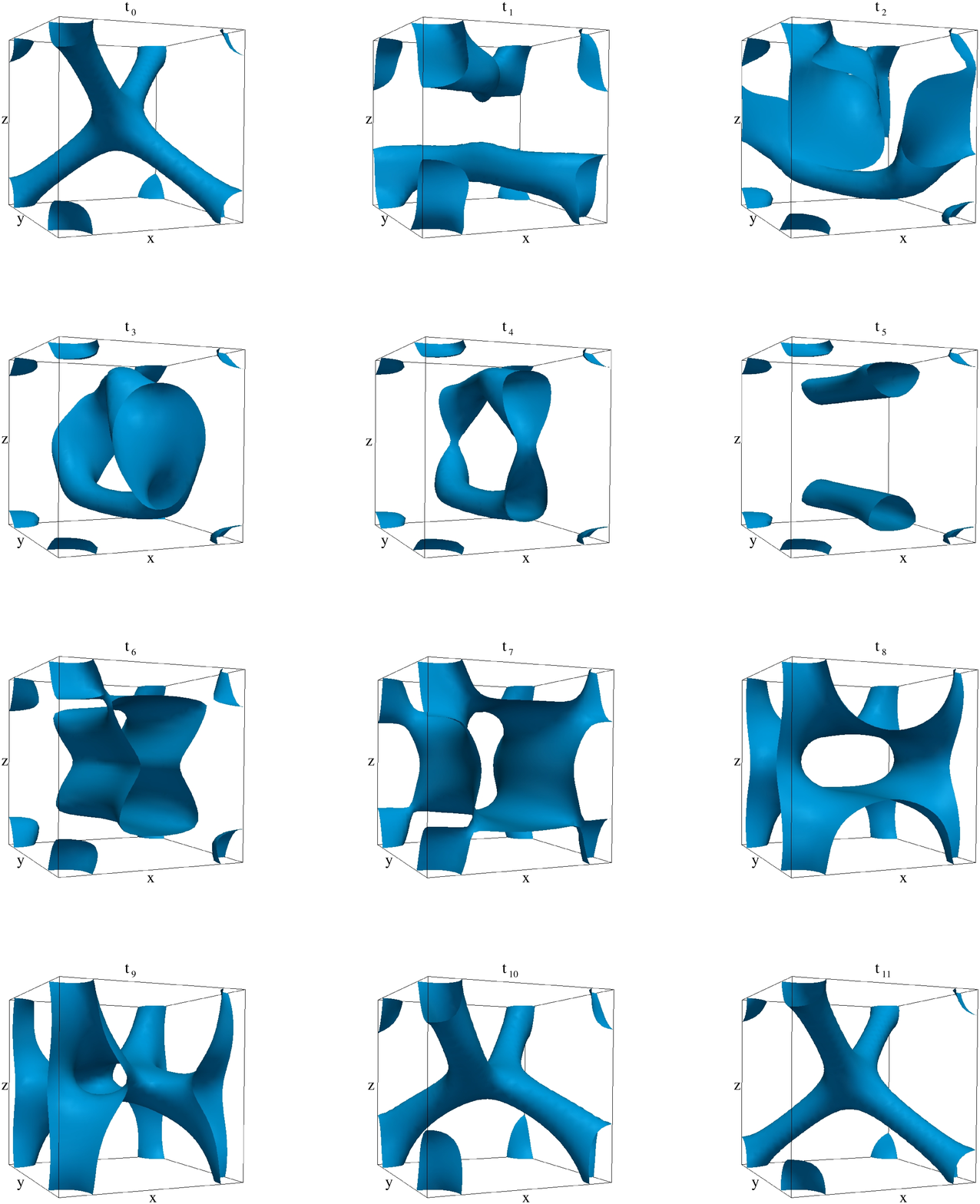}}
\caption{Evolution of the disclination network of the BPII phase during the switching on ($t_0-t_5$)
-off ($t_6-t_{11}$) dynamics in a cell of size $32\times 32\times 32$. 
Strong anchoring at the $z=0$ and $z=L$ planes are considered.
The electric field, applied along the $z$-direction ($[0,0,1]$), is switched on ($\Delta V = 0.4$) at $t_0=35000$ and
switched off at $t_5=150000$. After a cycle, the system reforms the original defect structure, as can be seen
at time $t_{11}$. For this reason the BPII cell can be used as a switchable device.
All the values reported are in simulation units.}
\label{fig3}
\end{figure*}

\begin{figure*}
\centerline{\includegraphics[width=17.5cm]{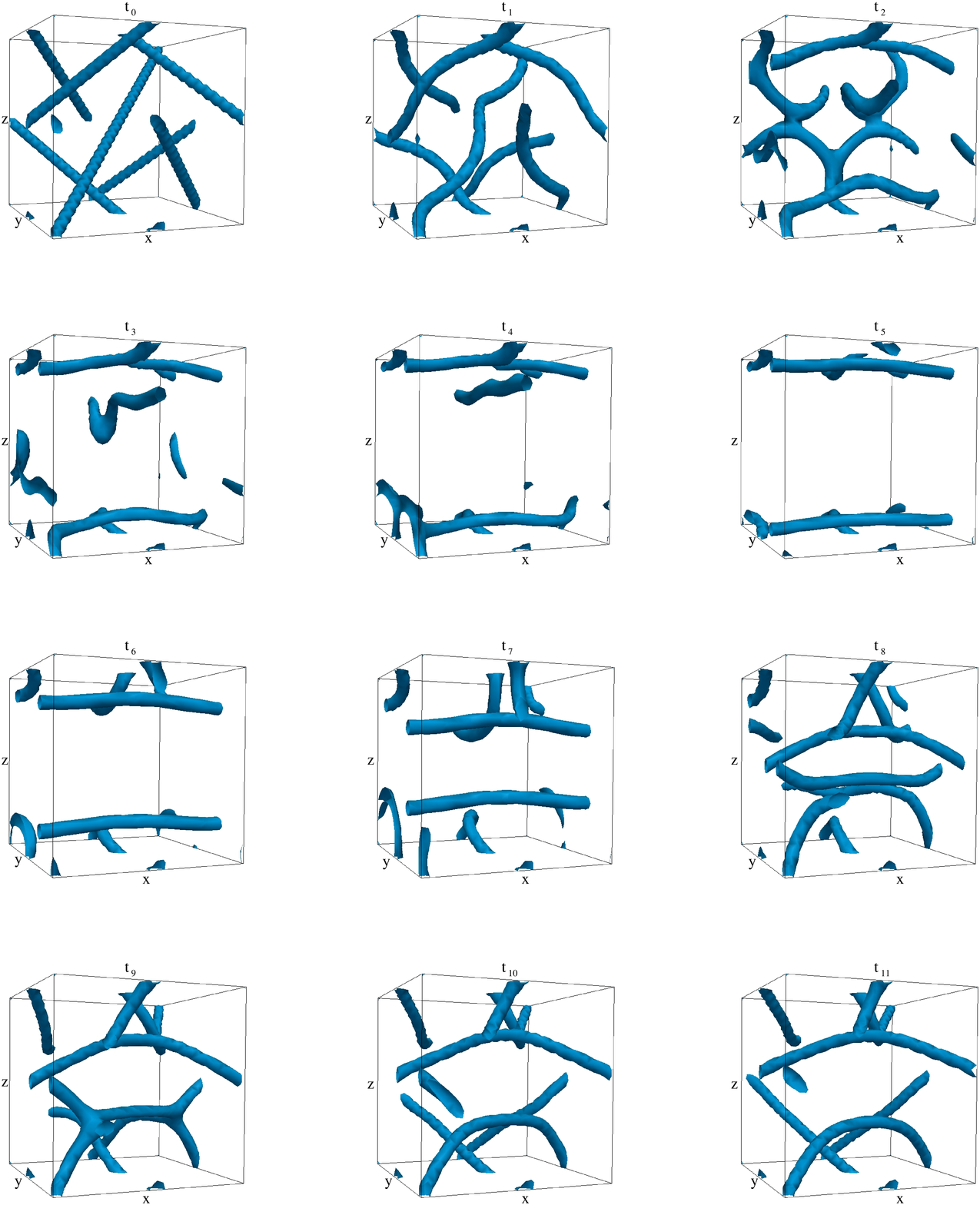}}
\caption{Evolution of the disclination network of the BPI phase during the switching on ($t_0-t_5$)
-off ($t_6-t_{11}$) dynamics in a cell of size $32\times 32\times 32$. 
Strong anchoring at the $z=0$ and $z=L$ planes are considered.
The electric field, applied along the $z$-direction ($[0,0,1]$), 
is switched on ($\Delta V = 0.5$) at $t_0=35000$ and switched off at $t_5=250000$. The final
state at time $t_{11}$ has a different defect structure respect to the typical network of the BPI
($t_0$). The free-energies of these two states are different (see Fig.~5) but, after several cycles,
the switching is reversible between the states $t_5$ and $t_{11}$.
All the values reported are in simulation units. }
\label{fig4}
\end{figure*}

Fig.~5 shows the time behaviour of the free energy for the BPII
(left) and the BPI (right) devices corresponding to the switching
on-off dynamics of Fig.~3 and 4. In both cases the electric field
was applied after $t_0=35000$ iterations, and correspondingly the
free energy abruptly drops. When the electric field is switched off
($t_5=150000$ (left) and $t_5=250000$ (right)), the free
energy increases abrubtly and then relax down to a new (zero field)
steady state. From the inset it is shown that whereas the BPII
device is able to get back to the zero field state after removal of
the field, the BPI cell gets trapped in a metastable configuration,
with free energy slightly larger than that of the stable BPI phase.
{The failure of the BPI device to come back to its original
equilibrium state seems to be quite generic since we observe the same phenomena 
for a starting state that belongs to a different region of the phase diagram 
(see squares in Fig. 5 right panel).}

\begin{figure*}
\centerline{\includegraphics[width=19.cm]{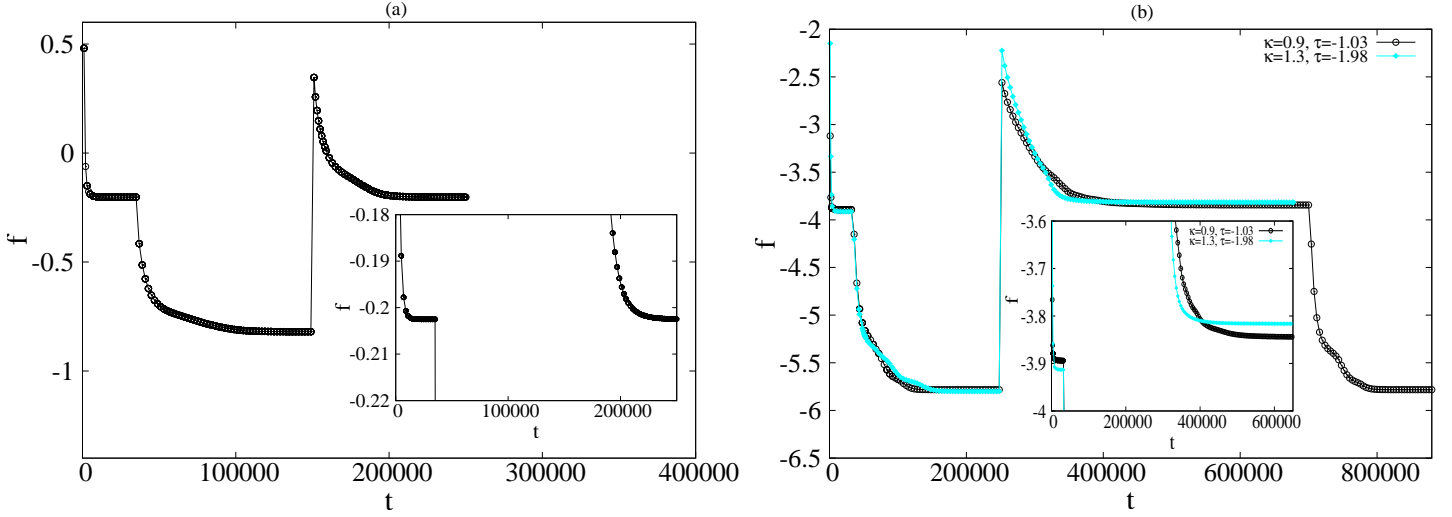}}
\caption{ Time evolution of the free energy of the BPII (left) and BPI (right) phases
during the switching on-off dynamics of Fig.~3 and 4. The two insets are zooms of the main
figures and show the values of the free energies at the beginning and at the end of the
switching on-off cycle. One can notice that, while for BPII the two values are essentially the
same, for the BPI phase the final state has a higher values of the free energy (metastable state). 
{
This turns out to be true even if we start from initial states belonging to 
different regions of the phase diagram (open circles and filled squares, see legend for values of $\kappa$ and $\tau$). Open circles correspond to the parameters in Fig. 4; filled squares to $A_0=0.0012$, $K=0.005$, $\gamma=5.085$, and other parameters as above.}}
\label{fig5}
\end{figure*}

As the blue phases are anisotropic phases, the orientation of the
electric field matters, and we have compared the case presented
above in which the field was along the $\hat{z}$ axes to the one in
which it is applied along the [111] diagonal of the unit cell. The
results are shown in Figs.~6 and 7, for BPII and BPI respectively.
{Note that here we consider the dynamics for values of the field which are sufficient to break the initial networks, so that these can be compared with the switching dynamics with the field in the [001] direction.}
The case of a diagonal electric field was previously considered in Ref.~\cite{fukuda2}, although in that work backflow effects were not implemented 
and  periodic boundary conditions were considered in
all directions.

In the case of BPII (Fig.~6), the switching on dynamics
is now different. In particular, the network now does not unbind,
and the central defect which joins up four disclinations is now
stable: it moves along the direction of the field until it touches
the surface to recombine into a pair of arcs ($t_0-t_5$). 
Once more, when the field
is removed the network grows from the surface into the bulk and
reconstructs the initial configuration ($t_6-t_{11}$).

\begin{figure*}
\centerline{\includegraphics[width=17.5cm]{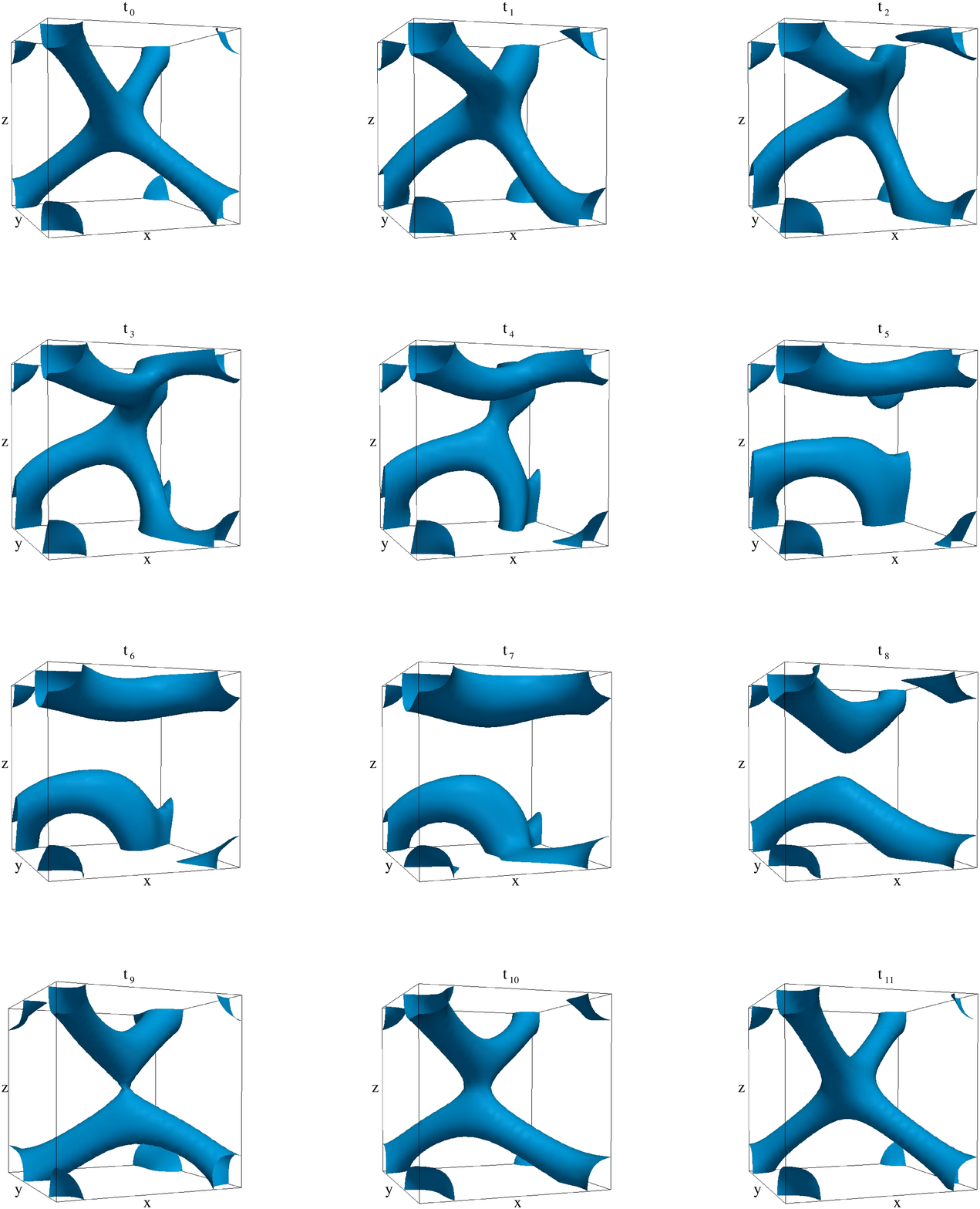}}
\caption{Evolution of the disclination network of the BPII phase during the switching on ($t_0-t_5$)
-off ($t_6-t_{11}$) dynamics in a cell of size $32\times 32\times 32$. Strong anchoring at the $z=0$ and $z=L$ planes are considered.
Unlike the case shown in Fig.~3 the electric field is switched on (at $t_0=35000$) along the
diagonal ([111], pointing vertically and to the right in the picture) of the unit cell
($\Delta V/L_x = \Delta V / L_y = \Delta V/ L_z = 0.2$). 
The switching off occurs at $t_6=305000$. As in the case in which the electric field
is applied along the $z$-direction, here, after a cycle, the network disclination ($t_{11}$) is
identical to zero-field configuration observed at time $t_0$.
All the values reported are in simulation units.}
\label{fig6}
\end{figure*}

Fig.~7 shows the switching of a BPI cell with a field along the
[111] directions. As for the case of the field along [001], the
disclination twists in an applied field -- remarkably, the one along
the field direction however does not ($t_1$). When the deformation
becomes large enough, the network forms branching points, 
which are unstable and
annihilate in the bulk leaving once more a state with arcs of
defects close to the boundaries ($t_2-t_5$) -- this time the state
is not symmetric, unlike the analogous case with the field along
[001].  The switching off proceeds
along a similar pathway as the one shown in Fig.~4, even though the
final state is yet different ($t_6-t_{11}$). 

\begin{figure*}
\centerline{\includegraphics[width=17.5cm]{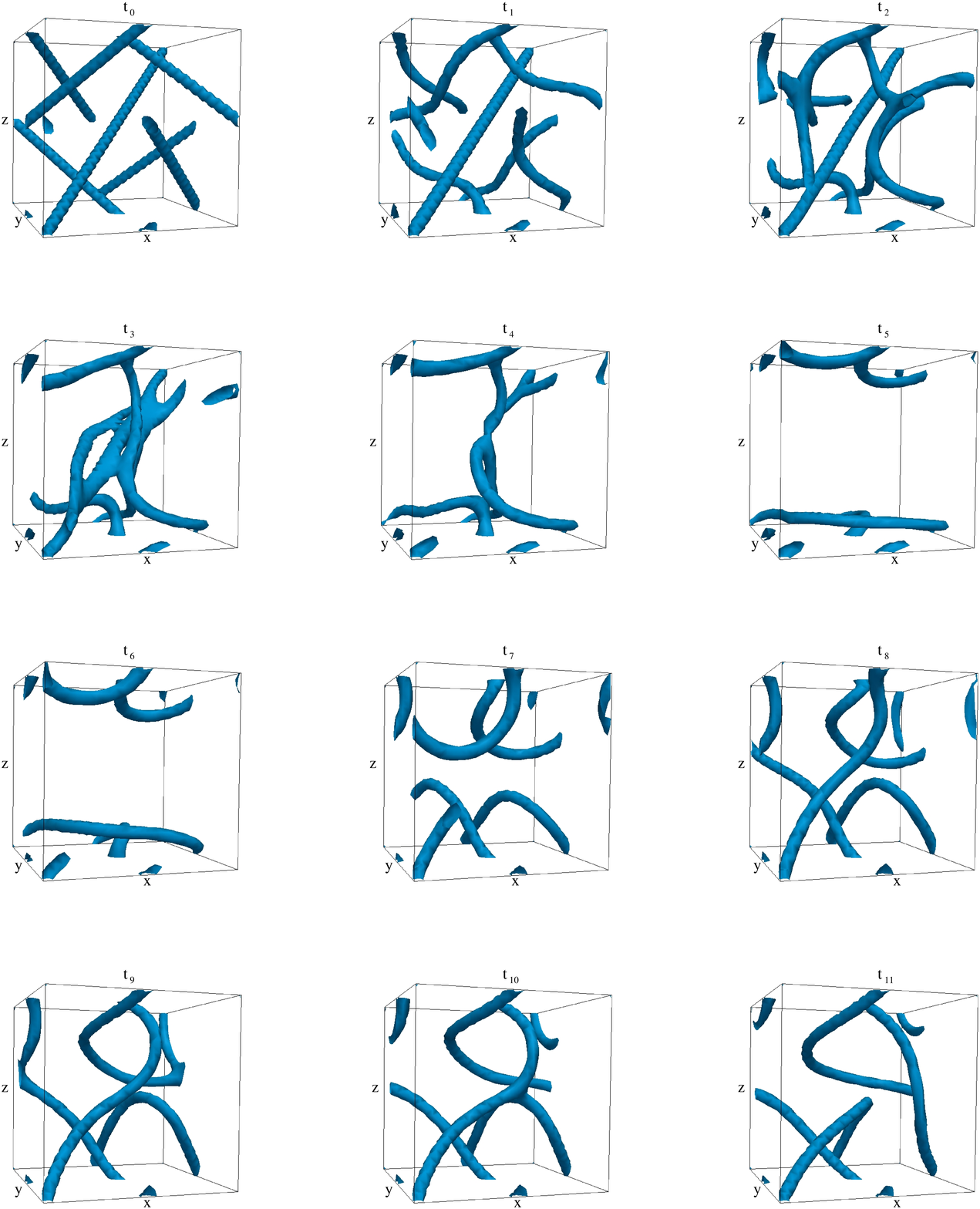}}
\caption{Evolution of the disclination network of the BPI phase during the switching on ($t_0-t_5$)
-off ($t_6-t_{11}$) dynamics in a cell of size $32\times 32\times 32$. Strong anchoring at the $z=0$ and $z=L$ planes are considered.
Unlike the case shown in Fig.~4 the electric field is switched on (at $t_0=35000$) along the
diagonal ([111]) of the unit cell
($\Delta V/L_x = \Delta V / L_y = \Delta V/ L_z = 0.275$).
The switching off occurs  at $t_5=250000$. Similarly to the case observed in Fig.~4, after a cycle,
the system can not recover the zero-field stable confioguration of the defects of the BPI.
All the values reported are in simulation units.}
\label{fig7}
\end{figure*}

In our simulations, we can switch off hydrodynamic interactions,
which is often referred to as backflow in the field of liquid
crystal dynamics, and therefore unambiguously pinpoint the effect of
hydrodynamics. Qualitatively, it is sometimes the case that
hydrodynamics allows the kinetics of the system to cross free energy
barriers which would otherwise trap it into metastable states.
Quantitatively, the relaxation times are usually decreased by
backflow, which ``speeds up'' many processes. 
 Fig.~8 shows the comparison between two
switching on and off simulations, one with hydrodynamics and the
other without -- we considered the case of a BPI cell with field
along [001]. It can be seen that
hydrodynamics slightly accelerates the dynamics. The steady states are not
affected for BPII; they are different for BPI, 
but are metastable in both cases and very close in free energy.

\begin{figure*}
\centerline{\includegraphics[width=8.cm]{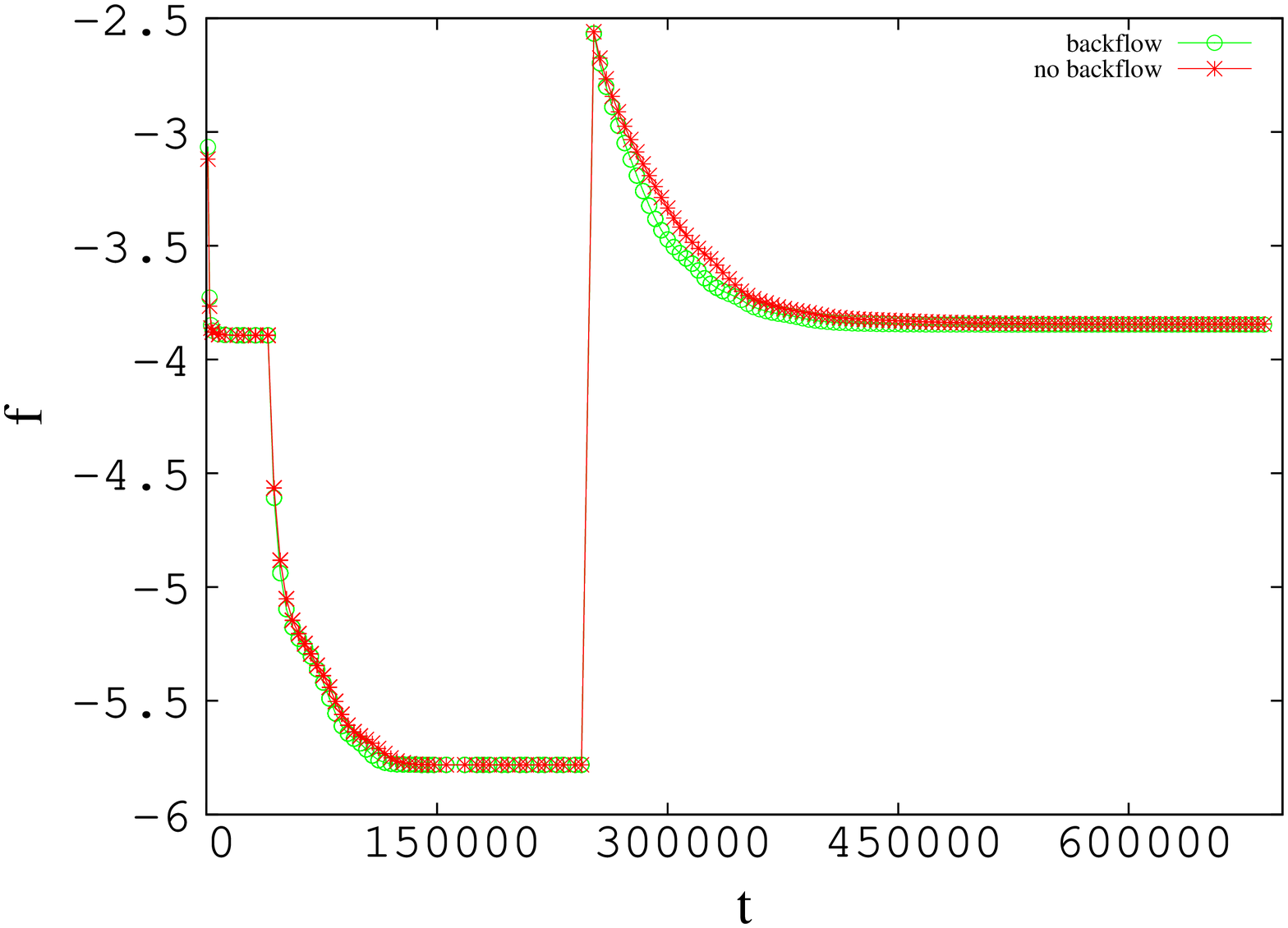}}
\caption{ Time evolution of the free energy of the BPI phase
during a switching on-off dynamics. 
The electric field is taken along the [001] direction.
The stars (circles) correspond to the situation withouth (with) hydrodynamics. }
\label{fig8}
\end{figure*}

\subsection{Effect of boundary conditions}

We now come back to the important, and non-trivial, issue of the choice of the boundary conditions for our device switching simulations. As mentioned above, fixed boundaries have often been used in the physics of cholesteric liquid crystals and blue phases~\cite{perm1,perm2,perm3,alex2}. Indeed, in the case of permeation, fixed boundaries have proved to be closer to rheology experiments than, say, the case in which the director field can rotate freely at the boundary~\cite{perm3}. Still, these boundary conditions are very difficult to exactly enforce in experiments. In practice, as for cholesterics~\cite{perm1,perm2}, one may envisage that the liquid crystalline texture and disclination pattern may be pinned at least partially by impurities, or it may be kept for some time due to surface ``memory'' effects which slow down the dynamics close to the boundaries. In the case of our blue phase device, for large electric field the regrowth of a disclination pattern after switching off (BPII or the metastable BPI state in Fig. 4) is likely to be efficient due to the imprinting at the boundaries. To see how realistic this is, one would need experiments like those in Ref.~\cite{psbp}, but for large electric fields which destroy the initial blue phase at the bulk. For blue phase devices such as those in~\cite{psbp} the switching back occurs on the sub-ms scale: if surfaces respond as free boundaries, however, there should be a field above which regrowth is much slower, as it should then occur via nucleation of the stable zero-field blue phase starting from a field-induced metastable state. New experiments would be needed to completely settle this issue.

On the other hand, experimentalists do have at
their disposal a range of well developed and established techniques
to accurately control the alignment of the director field at the
boundaries of a liquid crystalline sample or device. For instance,
by rubbing it is possible to obtain a homogeneous anchoring,
parallel to the boundary plane, whereas by means of chemical
treatment the surface may be made to favour homeotropic, or
perpendicular, anchoring. These two anchoring conditions are very
much used in practice and, as they are very well controllable, therefore it
would be desirable to know what their effect is on the director
conformation and switching of blue phase devices. Indeed, some
simulations with relatively small sample sizes and no hydrodynamics
suggested that at least when the boundaries are close by homeotropic
anchoring~\cite{fukuda1}, boundary conditions change the defect structure
of a confined BPI phase.

To address the role of homeotropic and homogeneous anchoring on the
structure of a BP device at zero applied field, we have numerically
simulated the relaxation to equilibrium of three BPI and BPII unit
cells, in a $32\times 32\times 96$ simulation domain 
{(the size was increased to minimize the effect of boundary layers).} 
Periodic boundary conditions were
applied along the $x$ and $y$ directions, whereas (infinitely)
strong homeotropic anchoring was enforced on the $z=0$ and $z=L$
planes. Fig.~9 shows the steady states for BPII (a) and BPI (b)
respectively. In both cases, while the bulk of the sample is
virtually indistinguishable from an equilibrated BPI/BPII cell,
there are strong anchoring driven deformations close to the
boundary. In the case of BPII (panel a) these result into the
formation of a stack of parallel disclinations confined at the
surface and disjointed from the rest of the network. In the BPI
cell, on the other hand, the diagonal disclinations approaching the
boundaries bend away from these, and again more disclination lines
appear at the surfaces, although this time they are not straight,
but twisted, and the lines on the two boundaries are perpendicular
to each other.

\begin{figure*}
\centerline{\includegraphics[width=13.cm]{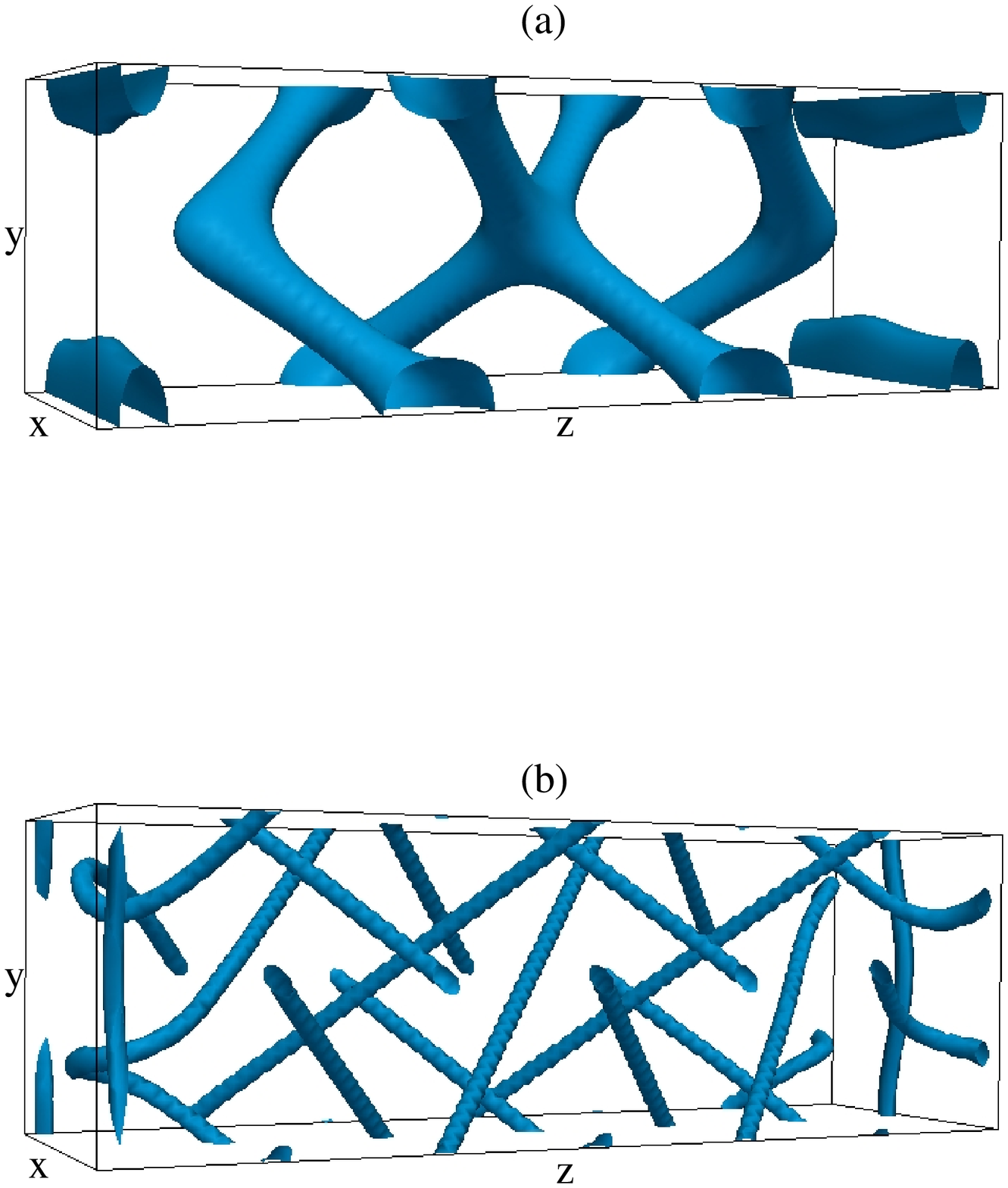}}
\caption{Steady state disclination networks at zero field for BPI (panel b) and BPII (panel a).
The simulation domain has size $32\times 32 \times 96$. Periodic boundary conditions are considered
along $x$ and $y$ whereas strong {\em homeotrophic} anchoring has been enforced on the planes
$z=0$ and $z=L$. The effects of boundaries are limited to the size of a single domain. The bulk
structure is unaffected while column-like surfaces appear near the boundaries.}
\label{fig9}
\end{figure*}

Fig.~10 shows the steady state configurations found with {\em homogeneous} anchoring, along the $z=0$ and $z=L$ planes.
In this case, there are no
surface defects disjointed from the bulk of the network, rather the
network deforms and reconstructs to join just above the surface and avoids it.
As in the homeotropic case, in our 3 cell domain the bulk is
basically unaffected by the anchoring, which suggests that surface
effects do not appreciably extend over one unit cell.

\begin{figure*}
\centerline{\includegraphics[width=13.cm]{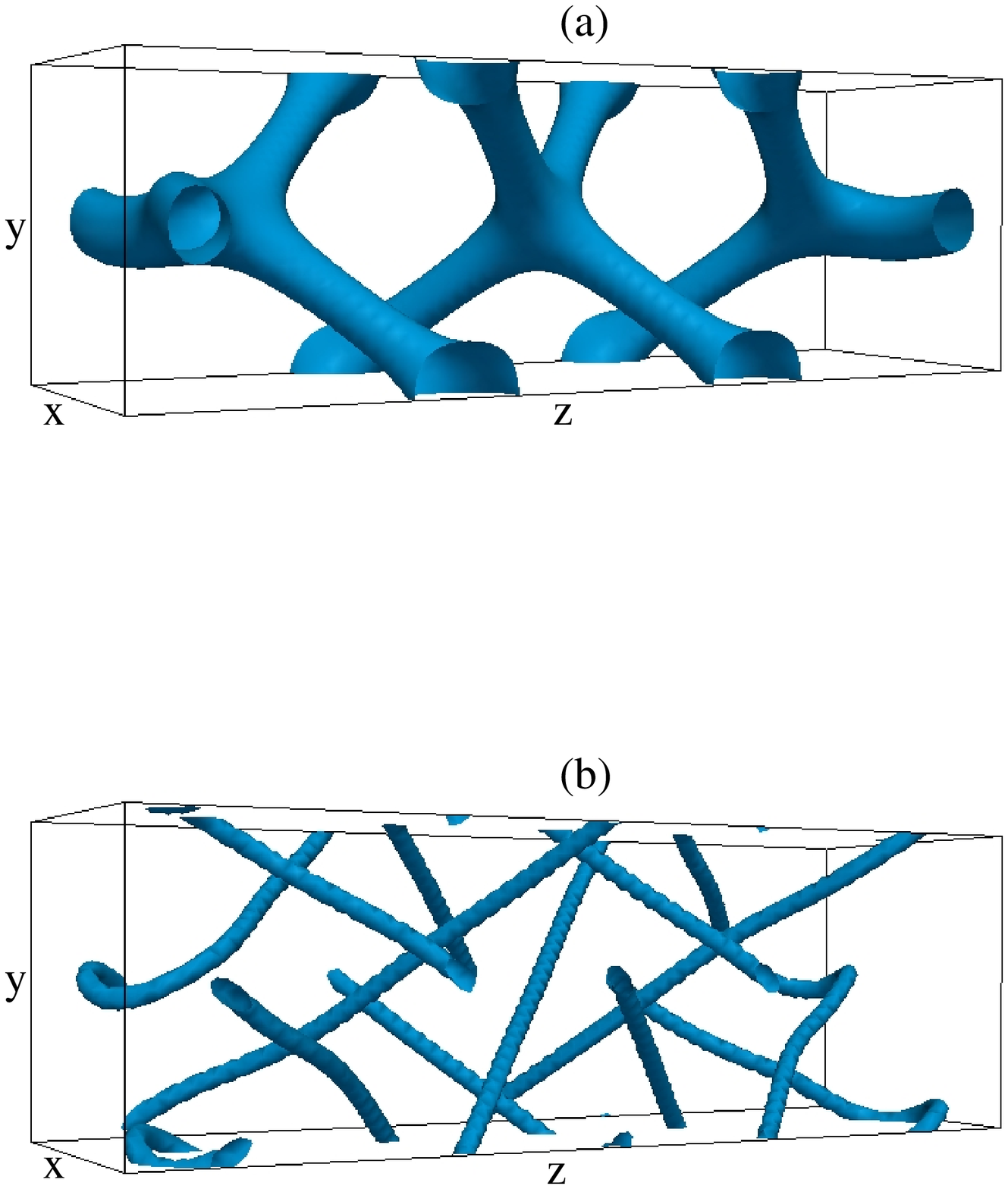}}
\caption{Steady state disclination networks at zero field for BPI (panel b) and BPII (panel a).
The simulation domain has size $32\times 32 \times 96$. Periodic boundary conditions are considered
along $x$ and $y$, and strong {\em homogeneous} anchoring along the $x$ direction
 has been enforced on the planes $z=0$ and $z=L$.}
\label{fig10}
\end{figure*}

{We have therefore seen that changing anchoring conditions has a strong effect on the zero-field structure of the network close to the surface. What about the behaviour in an electric field? Rather than doing an exhaustive analysis, which is too costly computationally, we have chosen here to focus on one case, that of BPI with homeotropic anchoring. This should be the most interesting case, as hysteresis effects and metastability are more pronounced for BPI. Fig.~11 shows the switching dynamics with $\Delta V=1$. 
After the field is switched on along the $\hat{z}$ axis, 
we observe that the bulk disclinations twist, as in the single unit cell simulation 
(see Fig.~1e), while the twisted defects appearing at the surfaces
reorganize as follows: some of them bend and move slightly towards the bulk while
the others remain fixed at the boundaries forming column-like structure
($t_0-t_2$). When the field is switched off ($t_3-t_5$) the bulk
network restructures itself, restoring the usual BPI defect disposition,
while at the surfaces only a partial restoration of the defect
structure as at $t_0$ occurs. Notice that for the voltage chosen here, 
$\Delta V=1$, in a device with a single unit cell, the
disclination network breaks and does not reform (Fig.~1). In the current simulations, however, permanent deformations in the defect structure are observed
at the steady state only at the surfaces, while the structure in the bulk is
unaffected. 
We have also performed further simulations with 6 unit cells in the field 
direction (with a unit cell resolved with 16 instead of 32 lattice points) 
and homogeneous and homeotropic anchoring. We have studied the case of a 
very large field which destroys the disclination networks.  In this case 
the network does not reform. This may be seen as a prediction as it is possible to treat 
boundaries in liquid crystals so as to force such alignments.
Our results also suggest that the
boundary conditions are not expected to affect 
the qualitative behaviour of
the devices under an electric field, at least up to intermediate
values of the electric fields.}

\begin{figure*}
\centerline{\includegraphics[width=20.cm]{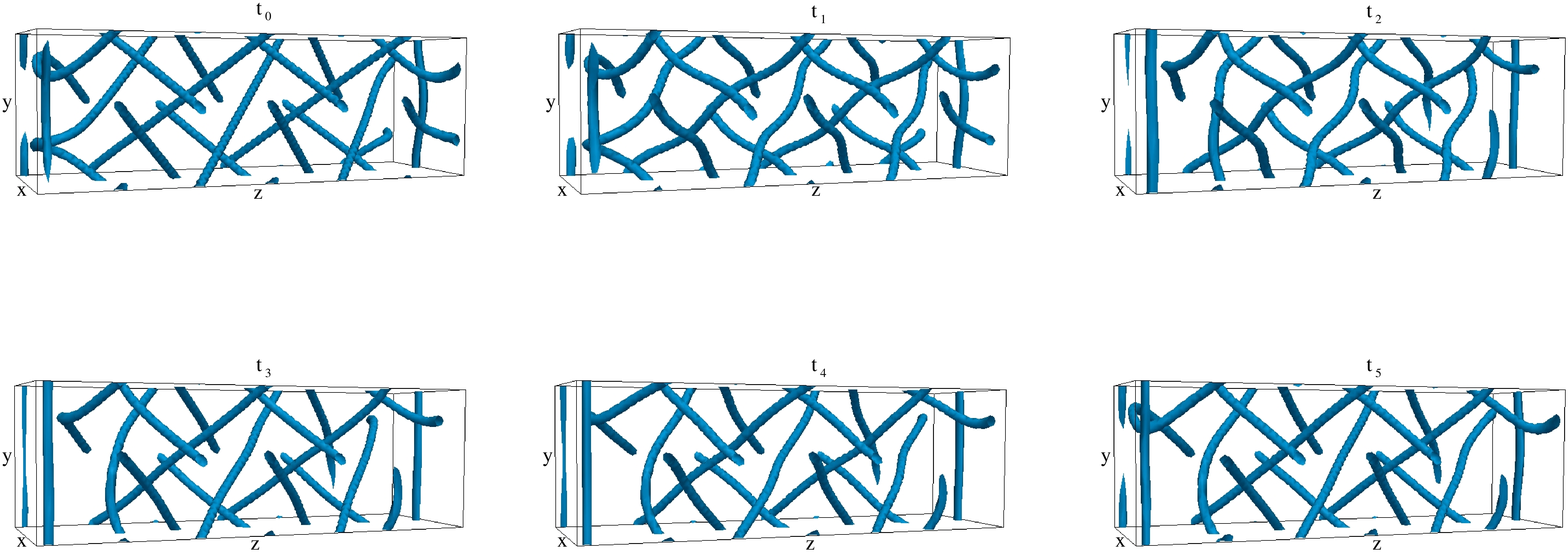}}
\caption{Evolution of the disclination network of the BPI phase during the switching on ($t_0-t_2$)
-off ($t_3-t_{5}$) dynamics. Homeotropic strong anchoring at the $z=0$ and $z=L$ planes are
considered.  The electric field is switched on ($\Delta V = 1.0$) at $t_0=120000$ along the
[001] (vertical) direction.
The field is then switched off at $t_3=376000$. }
\label{fig11}
\end{figure*}

\section{Conclusions}

In conclusion, we have reported a numerical investigation of electric field induced switching of devices built starting from cholesteric blue phases. At steady state, with a field along the [001] direction (which is normal to the boundary planes), BPII undergoes an \lq\lq unzipping transition\rq\rq upon increasing the electric field strength; while BPI twists above a first parameter dependent electric field threshold, then breaks as the field exceeds a second larger critical value. It is likely that the initial deformations of BPII (unzipping) and BPI (twisting) are crossovers or very smooth transitions, whereas the final defect disruption is a sharp transition.
The dynamics of the field-induced disruption of the network is very rich.

If the disclinations are fixed at the boundaries by impurities, our simulations predict an intricate pathway which allows the BPII structure to rotate its disclinations once they touch the top and bottom plates via the creation of a disclination loop. Once the field is switched off, we found that, remarkably, the liquid crystal can reform the original blue phase stable at zero field.
In the case of BPI, disclinations again twist up and then pin to the boundary. Upon switching off the field, the disclinations unpin and part of the network deforms, although the zero field configuration is not recovered unless the applied field is below a critical threshold. Given the vastly different defect and director configurations, the disclination dynamics predicted numerically should leave observable signatures in experiments.

When the field is along the diagonal of a unit cell, i.e. along [111], the BPII network does not unzip, rather it smoothly deforms until it touches the boundary. The disclinations in BPI, instead, still twist and reconstruct. The different response to the electric fields along [001] and [111] is a result of the anisotropy of blue phases. 

In practice, in liquid crystal samples the orientation of the director field at the boundary can be controlled via rubbing or chemical treatment. We have therefore investigated how our results are affected if the director field at the surfaces of the device is anchored and lies either perpendicularly to the surface or along one rubbing axis on the plane. In both cases, there is a boundary layer in which the disclination network reconstructs to either avoid the surface or form pinned disclinations disjoint to the rest of the network. If the device is one unit cell thick only, these deformations lead to new disclination networks, some of which are far from the original BP networks. For thicker devices, on the other hand, the bulk is essentially indistinguistable from the thermodynamically stable phase. In this regime, the switching in an electric field is not very different from the fixed surface boundary conditions. We also find that hydrodynamics, or backflow, has a relativel
 y minor effects on the switching on and off dynamics.

We hope that our results have shown that it is now possible to qualitatively follow and predict the switching on and off dynamics of BP devices with a variety of boundary conditions. The parameters we have chosen are similar to the typical liquid crystalline ones and
the effects observed are expected to be generic and robust for different values of e.g. viscosities and elastic constants.
It would also be interesting to study what happens if we relax the one elastic constant approximation alex''.  
Future experiments would be important to test the current simulation predictions and results, and to suggest further technologically relevant questions.

\vskip 1truecm

\noindent {Acknowledgements}

\noindent {We acknowledge the HPC-Europa2 program for funding a visit of AT to Edinburgh, during which this work was started, and CASPUR in Rome for a grant providing access to their computing facilities. We also warmly thank F. Salvadore, researcher at CASPUR, for help during the first stages of the work.}

\end{document}